\documentclass[11pt]{article}

\usepackage[final]{acl}

\usepackage{times}
\usepackage{soul}
\usepackage{latexsym}
\usepackage{multirow}
\usepackage{subcaption}
\usepackage{amsmath}
\usepackage{amssymb}

\usepackage[T1]{fontenc}

\usepackage[utf8]{inputenc}

\usepackage{microtype}

\usepackage{inconsolata}

\usepackage{graphicx}

\title{Do Assessment Instruments Measure the Same Thing for Humans and LLMs? A Latent Structure Analysis}

\author{
Alona Strugatski$^\ast$, 
\bf{Licol Zeinfeld$^\ast$}, 
\bf{Giora Alexandron}\\
\ Weizmann Institute of Science \\
{\small $^\ast$Equal contribution}\\[2pt]
\{alona.faktor, licol.zeinfeld,giora.alexandron@weizmann.ac.il\}\\
}

\begin{document}
\maketitle
\begin{abstract}
The rapid development and growing deployment of large language models (LLMs) have made it increasingly important to understand their capabilities. A common approach is to evaluate LLMs using assessment instruments originally designed to measure skills and competencies in humans, such as standardized exams, and to use performance on these instruments as evidence for generalizable claims about LLMs’ underlying abilities on the same skills the assessments are intended to measure in humans. However, from a validity perspective, such inferences require that the relationship between observed performance and underlying constructs established for humans also holds for LLMs. In particular, a necessary condition for transferring score interpretations is similarity in the latent structure of responses to the assessment. In this study, we examine whether this condition holds in two educational contexts: high-school chemistry and a quantitative reasoning section of a university entrance exam. Using a case study design, we compare human response data with responses generated by six multimodal LLMs. Our analytical approach combines exploratory factor analysis, factor congruence, and resampling to assess latent structure similarity across human learners and LLMs. Across both instruments, we find systematic differences between human and LLM factor structures, showing evidence that the analyzed assessments may not measure the same constructs for humans and LLMs. These findings call into question the validity of evaluation practices that use educational assessments to make claims about AI capabilities.
\end{abstract}

\section{Introduction}
The rapid development and growing deployment of large language models (LLMs) have made the systematic evaluation of their capabilities increasingly important \citep{raiaan2024review,laskar2024systematic}. Evaluating LLM performance across knowledge domains serves both as a benchmark of their progress and as a means of understanding how their capabilities compare to those of humans \citep{hendrycks2020measuring, zhu-etal-2025-tableeval}. Established assessment instruments, particularly standardized exams, are commonly used as evaluation frameworks \citep{achiam2023gpt, zhong-etal-2024-agieval}. The rationale is that using these instruments enables making generalizable claims about LLMs’ domain knowledge, skills, or competencies that the instruments were designed to measure \citep{sanchez-salido-etal-2025-bilingual, katz2024gpt, jimenez2023swe,borges2024could, Yacobson2026, chlapanis-etal-2025-greekbarbench}. 

This inference implicitly assumes that the link between test performance and underlying ability, validated in human populations, extends unchanged to AI systems. However, inferences from test performance to underlying skills depend on the validity of the assessment \textit{in context}, and validity established for a given human population does not imply that the same validity argument holds for LLMs  \citep{mitchell2026evaluating, pmlr-v267-wallach25a}. 

A key observation about validity is that it is not a property of the instrument, but of the interpretations and uses of scores, that is, the inferences drawn from observed performance on the instrument to the constructs it is intended to measure \citep{messick1994interplay}, which should be observed as probabilistic claims \citep{mislevy2003focus,liu-etal-2024-ecbd, xiao-etal-2023-evaluating-evaluation}. A necessary (although not sufficient) condition for transfer of an assessment instrument between populations for making the same interpretation is a similar underlying latent structure across the two populations, meaning that the instrument measures the same constructs \citep{meredith1993measurement}. Otherwise, the instrument simply measures different things.  The goal of the present study is to examine whether this condition holds in science education and quantitative reasoning contexts. It seeks to answer the following research question (RQ):  

\noindent\textbf{\textit{RQ}:} To what extent do assessment instruments in science education and mathematical reasoning exhibit similar latent structures when administered to humans and large language models?
Our primary analytical approach is based on exploratory factor analysis (EFA). EFA is the standard approach to analyze latent structure of assessment instruments, and similarity in factor structure can be interpreted as a preliminary indicator of construct equivalence across populations \citep{vandenberg2000review,meredith1993measurement}. Accordingly, substantial differences in the factor structures obtained from human and LLM responses would suggest that the instrument may not support the same construct-level interpretation across these populations. Here, we use it to examine whether similar latent structures emerge in human and LLMs response data on the same assessment instruments. 

We adopt a case study design, applying this analytical approach to two assessment instruments drawn from different STEM contexts: (1) a high-school chemistry diagnostic test completed by several hundred Grade 11–12 students, and (2) a standalone quantitative reasoning section from a high-stakes university entrance examination. The human datasets were augmented with responses generated by six multimodal LLMs (OpenAI: GPT-4o, GPT-5.2; Google: Gemini 1.5 Pro, Gemini 3 Pro; Anthropic: Claude 3.5 Sonnet, Claude 4.5).

The analysis proceeded in two steps. First, we applied EFA separately to human and LLM response data for each instrument, estimating the latent structure for each type of examinee and comparing the resulting structures qualitatively. Second, we compared the resulting structures using factor congruence, optimal factor matching, and repeated resampling to obtain a quantitative measure of structural similarity between LLMs and human responses. The results showed evidence that, across datasets and choices of the number of factors, LLM--human factor structures differ. 

The contribution of this work is both methodological and empirical: it introduces a validity-oriented framework for comparing the latent structure of assessments across humans and LLMs, and demonstrates, through two case studies, that established assessment instruments may capture substantially different constructs in humans and LLMs.

\section{Related Work}
\subsection{LLM Evaluation using Educational Assessments}
Assessment-based evaluation of LLMs has rapidly become common practice across domains, often by reusing instruments originally developed to measure human knowledge or reasoning \cite{Kasagga2025MedicalLicensingMeta,Balunovic2025MathArena,zhong-etal-2024-agieval,Du2025SuperGPQA}. In medicine, models are evaluated on medical exam-style benchmarks and clinical reasoning datasets \cite{Kasagga2025MedicalLicensingMeta,Alaa2025ConstructValidity}. In mathematics and science, they are tested on human-targeted problem-solving benchmarks such as AIME 2025 and domain-specific assessments in chemistry and materials science \cite{Balunovic2025MathArena,Zaki2024MaScQA,Arora2023HaveLLMsAdvancedEnough,Yacobson2026, Chen2025ScholarChemQA, Guo2023ChemLLMBench, Alampara2024MaCBench}. A similar pattern appears in law and general academic reasoning, where benchmarks such as AGIEval and SuperGPQA draw on standardized exams including the SAT, LSAT, law-related assessments, and graduate-level disciplinary questions \cite{zhong-etal-2024-agieval,Du2025SuperGPQA}. This growing practice reflects an implicit assumption that performance on human-designed assessments can be interpreted as evidence of LLM ability in the corresponding domain. This extends across benchmarks that differ in language, task modality, and task domain (such as sequential reasoning, higher-order cognitive tasks), including evaluations specifically constructed to probe current limitations of LLMs \cite{Ahuja2024Megaverse,Chang2024SurveyLLMEvaluation,Balunovic2025MathArena, chlapanis-etal-2025-greekbarbench}.

While evaluating LLMs via human-targeted assessments has become common practice, recent literature reveals limitations in this approach both generally and across specific domains \citep{liu-etal-2024-ecbd, xiao-etal-2023-evaluating-evaluation}. For example, in medicine, recent critiques highlight profound gaps in the validity of applying clinical exams to LLMs, noting that high scores on the MedQA benchmark fail to translate to actual clinical decisions relying on the same clinical knowledge \citep{Kasagga2025MedicalLicensingMeta}. Similarly, in the chemical sciences, established benchmarks like MoleculeNet \citep{wu2018moleculenet} are increasingly criticized for their narrow scope and their limited ability to provide insights into how models compare relative to human experts, particularly when dealing with specialized molecular structures or equations \citep{naturechem2025}. These domain-specific failures are often symptoms of deeper technical issues, such as \textit{data contamination} \citep{kapoor2023leakage, xu-etal-2025-contamination, sainz-etal-2023-nlp} or the \textit{brittleness} of model reasoning under minor task perturbations \cite{ullman2023large, inger2025forget} 

Beyond these domain-specific failures in scientific contexts, broader methodological concerns persist regarding current benchmarking practices. First, the field's heavy reliance on aggregate metrics often obscures how a system will perform in a specific situation, making it difficult to predict real-world performance. This issue is compounded by a lack of transparency; the instance-by-instance evaluation results necessary to ``unpack'' these aggregate scores are rarely made available for independent audit \citep{burnell2023rethinking}. More fundamentally, there is a growing recognition that while the capabilities of language models are advancing rapidly, the theoretical foundations and norms for their rigorous evaluation lag behind \citep{balepur2025bestdescribesmultiplechoice}, suggesting that established theories of educational measurement may provide a foundation for evaluating the cognitive abilities of artificial `intelligence' as well \citep{salaudeen2025validityframework,mitchell2026evaluating}.Recently, both theoretical interest and the empirical application of these ideas have been rapidly growing in the literature. Aligning with this expanding body of work, we empirically demonstrate evidence highlighting the limitations of making the same inferences about LLMs based on assessments developed for human learners, while also demonstrating the usefulness of established educational measurement workflows for developing methodologically grounded approaches to evaluating LLM capabilities.

\subsection{Human \& LLM Divergences on Educational Assessments} There is increasing evidence that LLMs do not behave like human examinees on educational assessments \citep{Yacobson2026}, and their responses often fail to exhibit consistent psychometrically plausible profiles \citep{petrov2024limited,strugatski2025irt}. For example, \citep{sauberli-etal-2025-llms} found that some models can be made somewhat more human-like through calibration, but the overall correspondence between LLM and human responses remains limited. Similarly, \citep{liu2025leveraging} shows that no single LLM adequately mimics human respondents, largely because model response distributions are too narrow, even when some psychometric properties can be approximated. In a related multiple-choice setting, \citep{sorenson2024identifying} demonstrated systematic differences between human and GenAI response patterns. In STEM education, it was reported that  ChatGPT was limited in its ability to solve engineering questions involving figures or diagrams \citep{borges2024could}, and struggled when required to make assumptions about the real world or solve under-specified problems in physics \citep{wang2024examining}. Similar observations that LLMs were impacted differently by task dimensions were also made in chemistry and biology education contexts \citep{watts2023comparing}, and such differences can be utilized by Differential Item Functioning methods to flag items that operate differently for humans and LLMs \citep{zeinfeld2026identifying}. Together, these studies provide consistent evidence of substantial divergence between human and LLM respondents in educational settings and motivate closer examination of assessment validity.


\section{Methodology}
\subsection{Overview}
Our primary analytical approach is based on exploratory factor analysis (EFA) in accordance with \citep{CUDECK2000265EFA}. For each instrument, we applied EFA separately to human and LLMs responses. 
The analysis pipeline comprised four stages: (1) preprocessing the data, (2) estimating the number of factors to retain using two factor-retention criteria, (3) fitting EFA models and extracting factor structures separately for humans and LLMs, and (4) comparing the resulting structures using factor congruence, optimal factor matching, and repeated sampling. The final stage enabled both LLMs--human comparisons and a human--human similarity baseline.
The full code, sample data and prompts to reproduce the analyses in the paper can be found in the
GitHub repository: (\href{https://github.com/LicolZeinfeld/latent-structure-analysis/}{\textcolor{blue}{\underline{link}}})


\subsection{Instruments \& Human Data}
\label{sec: repreducibility}
For the present study, we used response data from two different assessment settings. The first was a high-school chemistry diagnostic test administered through Moodle as preparation for the matriculation exam; it included 22 multiple-choice items and responses from 931 students (\(M = 71.49\), \(SD = 16.95\), out of 100). The second was the quantitative reasoning section of a national university entrance exam, consisting of 20 multiple-choice items and responses from over 4,800 examinees (\(M = 12.45\), \(SD = 3.75\), out of 20). Both instruments were multimodal, including text-only items as well as items with figures, images, or formulas. Responses in both datasets were coded dichotomously, with correct answers marked as 1 and incorrect answers as 0.

In navigating the inherent trade-off between data quality and reproducibility, we opted to use a non-public dataset. While this choice limits reproducibility, it was necessary to ensure data quality across two dimensions: (1) It ensured high-quality human response data from real learners making genuine effort under authentic conditions. (2) It increased confidence that the instrument and its solutions had not been previously exposed, which is a prerequisite for EFA comparison.
An additional advantage, beyond the scope of this paper, is that these context-specific materials  enable future domain-expert interpretation of the results.

\subsection{Collection of LLM Responses}
To generate LLM response data, we collected answers from six multimodal LLMs spanning three model families: OpenAI (GPT-4o, GPT-5.2), Google (Gemini 1.5 Pro, Gemini 3 Pro), and Anthropic (Claude 3.5 Sonnet, Claude 4.5). Since the assessments included figures, formulas, and other visual content, multimodal input support was required for all models. For each model and instrument, we collected 20 independent response sets, yielding 120 total responses across the six models per instrument (chemistry: M = 76.14, SD = 15.41 \& quantitative reasoning: M = 11.94, SD = 4.52). Responses were collected through the models' online user interfaces, which is the practical setting in which these large proprietary models are typically used. For each run, we initiated a new temporary chat session to reduce possible carryover from prior prompts.
For the primary analyses, responses were pooled across the six models to form a single LLMs group. Conceptually, pooling aligned with our research question, which concerns a comparison of latent structures for human respondents and LLMs as a class, rather than any individual model. Statistically, pooling increased variation in LLM response patterns, which is necessary for estimating the item pair correlations underlying EFA. After pooling, the SD of LLM response patterns was comparable to that of the humans'. This pooling approach is consistent with recent NLP evaluation work \citep{macko-etal-2023-multitude}. Additionally, pooling across LLM models is consistent with how EFA estimates aggregate response structure rather than assuming homogeneous human respondents.

\subsection{Prompting Technique \& Response Scoring}
For each instrument, the full instrument was uploaded as a PDF through the LLMs' web interface together with an instruction asking the models to provide only the final answer choice for each item (complete prompt: (\href{https://github.com/LicolZeinfeld/latent-structure-analysis/}{\textcolor{blue}{\underline{link}}})). This was done to simulate a realistic student-exam setting, where students are presented with the full exam.
We used a minimal, zero-shot prompting approach, similar to \citep{munker-2025-fingerprinting}, aiming to elicit direct responses and avoid specific outcome optimization.
LLM behavior can be sensitive to prompt design \citep{zhuo-etal-2024-prosa}, and in our case, that would introduce an uncontrolled source of variation to the responses, which would make it harder to attribute response patterns to the assessment itself.
Model outputs were binarized against the answer key, and skipped or invalid responses were scored as incorrect, consistent with the treatment of human respondents.
\subsection{Experimental Flow}

\subsubsection{Preprocessing}
\label{subsec:preproc}
Given the binary item-response matrices, we used tetrachoric correlations, which estimate item associations by treating observed correct/incorrect responses as binary indicators of underlying continuous response tendencies. These correlation matrices served as input to all factor-retention and EFA procedures. As a preprocessing step, we screened items that could make these matrices unstable, particularly in the LLM response data. Specifically, we removed items with zero variance in either group and inspected the \(2 \times 2\) contingency tables used to estimate each pairwise tetrachoric correlation, since zero cells or very small cell counts can lead to unstable estimates.
Based on these diagnostics, items contributing to highly sparse pairwise tables were removed until the remaining item set reached an acceptable stability threshold. This preprocessing step resulted in no item removals for the quantitative reasoning item set, whereas seven items were removed from the chemistry item set: Items 1, 8-10, 12, 17, and 20. Analyses were then restricted to the common set of remaining items across humans and LLMs, ensuring both groups were analyzed on the same item set. 

\subsubsection{Factor-retaining}
\label{subsubsec:FacRet}
Before fitting EFA models, we estimated the number of factors to retain using two common retention methods: the Kaiser criterion and Parallel Analysis \citep{Sunbok2017FA,Nazaretsky2022Trust}. Both methods were applied separately to the preprocessed human and LLM tetrachoric correlation matrices for each instrument. This allowed us to examine whether humans and LLMs showed systematically similar or different evidence regarding the number of underlying factors. Differences at the factor-retention stage are already informative, as they suggest that the assessment may not reflect the same underlying structure across human and LLM responders. Comparing across factor-retention methods allowed us to test (1) between-group robustness: whether observed LLM--human differences in latent structure were method-dependent or robust, and (2) within-dataset stability: whether the item-loading pattern within each dataset remained stable across the different factor-retention methods.

\textbf{Kaiser Criterion}
This method retains factors with eigenvalues greater than one \citep{Kaiser1974}. When factor extraction is based on a correlation matrix, each standardized observed variable contributes one unit of variance. Therefore, an eigenvalue greater than one indicates that the factor explains more variance than a single observed variable. The Kaiser criterion is often used alongside other retention methods, such as parallel analysis.

\textbf{Parallel Analysis}
We also applied parallel analysis as a factor-retention method \citep{Timmerman2011Parallel}. Observed eigenvalues were compared to those obtained from 30 simulated datasets with the same numbers of items and respondents using the \texttt{fa.parallel} function from the \texttt{psych} package in R \cite{revelle2025psych}. Factors were retained as long as the observed eigenvalues exceeded those expected under random data. 
Because parallel analysis may yield different factor-retention results across repeated runs, we report the retained number of factors together with its stability, defined as the frequency with which the same number of factors was recovered within each run (see Subsection 4.1).

\subsubsection{Factor-extraction}
\label{sec:factor-extraction}
Next, EFA models were fit separately for the human and LLM groups based on the number of factors determined in the factor-retention step, in order to estimate the loading of each item on each factor. We used the default least squares optimization (psych:fa method: \texttt{fm = "minres"}) to minimize the difference between the observed and model-reproduced correlation matrices. We also enabled correlation between factor solutions using \texttt{rotate = "oblimin"} \cite{revelle2025psych}. 

\subsection{Factor Structure Similarity: LLMs vs. Humans} 
\label{subsec:factor-similarity-analysis}
To quantify the retention of latent factor structure within humans and compare it to LLM-human similarity, we conducted a repeated resampling analysis on the two binary-response datasets --  the chemistry and the quantitative reasoning. For each dataset, analyses were run separately with the number of factors fixed to 4, 5, 7 and 8 that were found in \ref{subsubsec:FacRet}. In each iteration, we drew two independent samples of 120 human respondents, denoted $H_1$ and $H_2$, and one independent sample of 120 LLM respondents, denoted $B$. We set the sample size to 120 because this was the number of available LLM response sets, allowing the human and LLM comparisons to be conducted under matched sample-size conditions. 

For each resampled subset, we fit EFA models following the procedure in \ref{subsec:preproc}. This yielded loading matrices for $H_1$, $H_2$, and $B$. We first established a human baseline by comparing the factor structures of $H_1$ and $H_2$. We then assessed LLM-human similarity by comparing the factor structure of $B$ to that of $H_2$ from the same iteration.

Let $\mathbf{A}, \mathbf{B} \in \mathbb{R}^{p \times k}$ denote two factor-loading matrices defined on the same $p$ items, each with $k$ factors. For factor $r$ in $\mathbf{A}$ and factor $s$ in $\mathbf{B}$, let $\mathbf{a}_r$ and $\mathbf{b}_s$ denote their corresponding loading vectors across items. Factor congruence \citep{Tucker1951AMF} was computed as the cosine similarity between these two loading vectors:
\begin{center}
    \(\phi_{rs} = \frac{\mathbf{a}_r^\top \mathbf{b}_s}{\|\mathbf{a}_r\|\,\|\mathbf{b}_s\|}\)
\end{center}


This yields a congruence matrix $\mathbf{\Phi} \in \mathbb{R}^{k \times k}$, where the $(r,s)$ entry is $\phi_{rs}$. Because factor order is arbitrary across EFA solutions, we matched factors using the Hungarian algorithm \citep{Kuhn1955Hungarian}, a loss-based linear assignment method that finds the optimal one-to-one correspondence between factors across the two loading matrices. Matching was based on the absolute values of the factor congruence coefficients, so that factors with the same structure but opposite sign were treated as equivalent. We used \(1 - |\phi_{rs}|\) as the matching loss, allowing the Hungarian algorithm to maximize absolute congruence through its standard minimization formulation. Each comparison was then summarized by the mean absolute congruence across the matched factor pairs.

This procedure was repeated 100 times for each subset and each factor-number choice, yielding distributions of mean matched congruence scores for the human-human (HH) and LLMs-human (LH) conditions. 
We compared the HH and LH conditions using one-sided Wilcoxon rank-sum tests across iterations, testing the null hypothesis that HH matched congruence was less than or equal to LH matched congruence against the alternative that it was greater.


\begin{figure*}[t]
  \centering
  \begin{subfigure}[t]{\linewidth}
    \centering
    \includegraphics[width=0.7\linewidth]{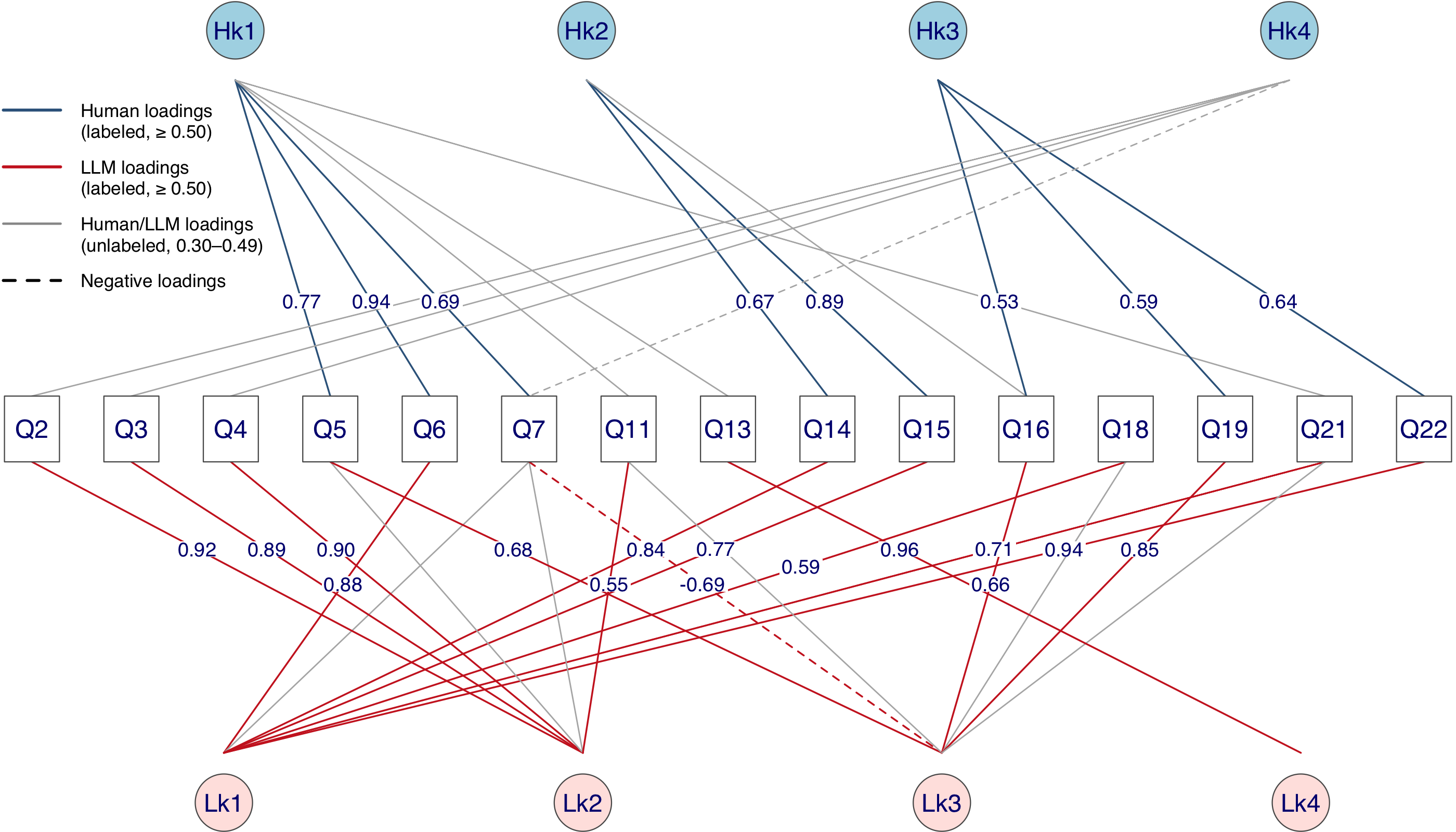}
    \caption{EFA structure of the Chemistry instrument for human responses (blue) and LLMs responses (red).}
    \label{fig:chem_kaiser_EFA_structure}
  \end{subfigure}

  \vspace{1.75em}

  \begin{subfigure}[t]{\linewidth}
    \centering
    \includegraphics[width=0.7\linewidth]{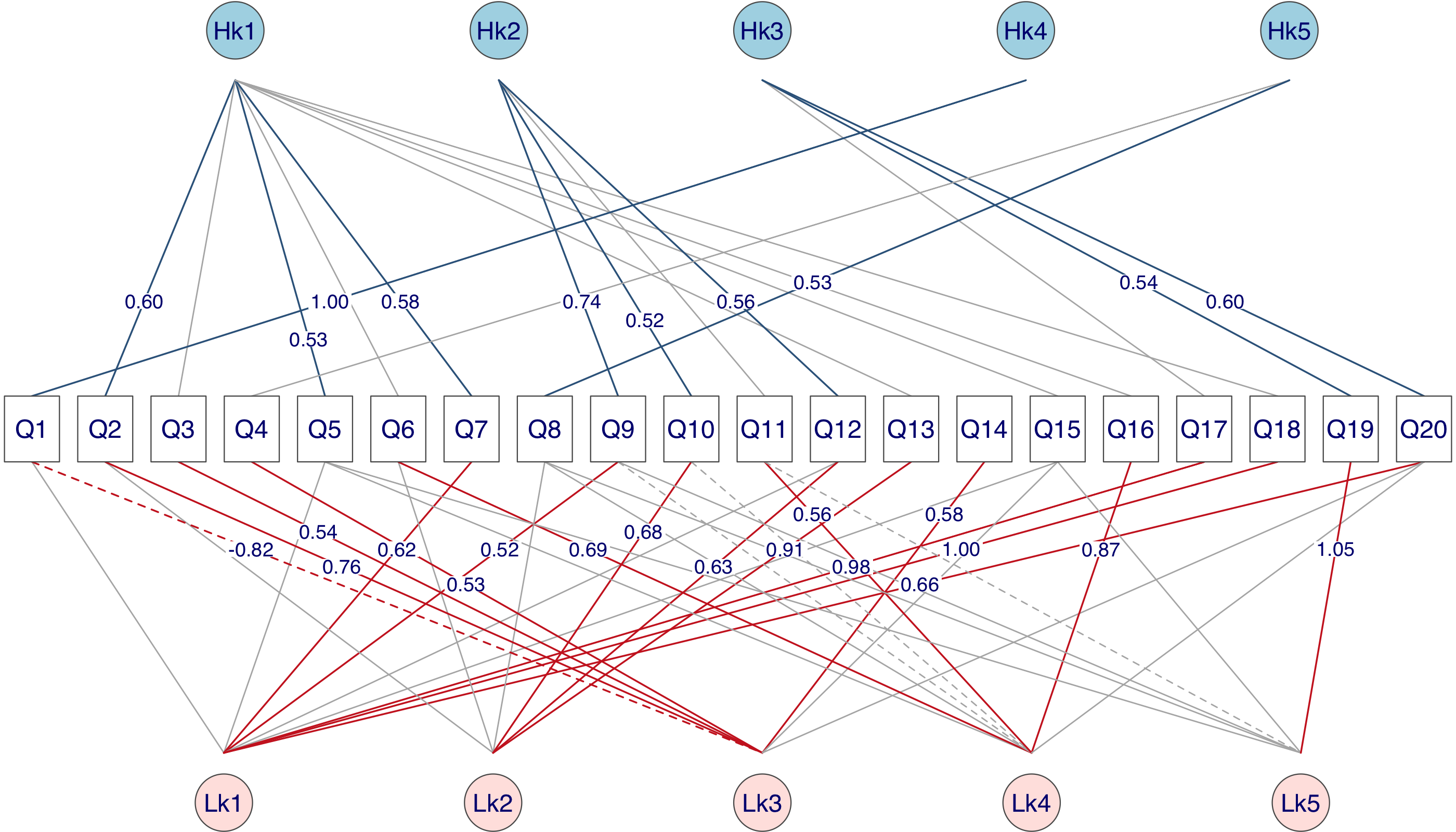}
    \caption{EFA structure of the Quantitative Reasoning instrument for human responses (blue) and LLMs responses (red).}
    
    \label{fig:psych_kaiser_EFA_structure}
  \end{subfigure}

  \caption{\textbf{Human vs. LLM EFA Structures Across Instruments.} Factor retention based on Kaiser criterion.} 
  \label{fig:kaiser_EFA}
\end{figure*}

\section{Results}
\subsection{Factor-retention}
Table~\ref{tab:factor_retention_results} summarizes the factor-retention results for both instruments under the Kaiser criterion and parallel analysis. The Kaiser criterion yielded matching retention results within each instrument -- four for the Chemistry, and five for Quantitative reasoning across the two groups. However, parallel analysis yielded different factors between the humans and the LLMs groups in both datasets. In Chemistry, humans consistently retained five factors, whereas LLMs most often retained 4. In Quantitative Reasoning, humans retained 7--8 factors, whereas LLMs consistently retained 5 (same as kaiser retention). With respect to differences in factor retention between groups, the results were mixed and depended on the retention method: the Kaiser criterion yielded the same number of factors for both groups, whereas parallel analysis yielded different numbers.    

\subsection{Factor-extraction}
Figure~\ref{fig:kaiser_EFA} shows the EFA factor-loading structures obtained for humans and LLMs for both instruments under the Kaiser-retained factor solution. As seen in the figure, although the retained number of factors was similar for humans and LLMs (4 for Chemistry with human \texttt{RMSEA} = 0.019 \& LLM \texttt{RMSEA} = 0.045, 5 for Quantitative Reasoning with human \texttt{RMSEA} = 0.011 \& LLM \texttt{RMSEA} = 0.092; for all, confidence \> 0.90), the loading structure is different. For instance for Chemistry, the third LLM factor, Lk3 (L: LLMs, K: Kaiser, Factor: 3), has loadings larger than 0.5 on Q5, Q16, and Q19. On the human side, these items are loaded on two different factors: Hk1 and Hk2. A similar thing happens with the factor structures of the two groups for the Quantitative Reasoning instrument. One example, among several, is Lk1, which has loadings greater than 0.5 on Q7, Q9, Q17, Q18, and Q20. On the human side, these items are loaded on three different factors: Hk1, Hk2, and Hk3. Analyzing the EFA with parallel analysis-retained factor number also revealed dissimilarities in the item-factor structure and loadings for both instruments (results figure was omitted due to space limit.)
To conclude, these factor analyses qualitatively demonstrate differences in the latent structure of the instruments for humans and LLMs. In the next section, we turn to examining these differences quantitatively using factor congruence analysis. 

\begin{table*}[t]
\centering
\small
\setlength{\tabcolsep}{6pt}
\renewcommand{\arraystretch}{1.12}
\begin{tabular}{@{}l|cc@{\hspace{12pt}}|@{\hspace{12pt}}cc@{}}
\hline
\multirow{2}{*}{\textbf{Instrument}} 
& \multicolumn{2}{c@{\hspace{12pt}}|@{\hspace{12pt}}}{\textbf{Kaiser Criterion}} 
& \multicolumn{2}{c}{\hspace*{4pt}\textbf{Parallel Analysis}} \\[4pt]
& \textbf{Humans} & \textbf{LLMs} & \textbf{Humans} & \textbf{LLMs} \\
\hline
Chemistry & 4 & 4 & \multicolumn{1}{l}{\textbf{5: 100\%}} & \multicolumn{1}{l}{\textbf{4: 62.4\%}, 3: 16\%, 5: 16\%, 2: 5.2\%, 7: 0.4\%} \\
Quantitative Reasoning & 5 & 5 & \multicolumn{1}{l}{\textbf{8: 66\%}, 7: 34\%} & \multicolumn{1}{l}{\textbf{5: 100\%}} \\
\hline
\end{tabular}

\caption{Factor-retention results for the Chemistry and Quantitative Reasoning instruments under the Kaiser criterion and parallel analysis. In the parallel analysis, the percentages refer to the proportion of runs that yielded each result.}
\label{tab:factor_retention_results}
\end{table*}


\subsection{LLM–Human Similarity Analysis }

\begin{table}
\centering
\small
\begin{tabular}{llccccc}
\hline
\textbf{Inst.} & \textbf{$k$} & \textbf{L-H sim.} & \textbf{H-H sim.} & \textbf{Cohen's $d$} & \textbf{$W$} \\
\hline
\multirow{4}{*}{Quant.} 
& 4 & 0.438 & 0.543 & $1.80^{***}$ & 8987.0 \\
& 5 & 0.432 & 0.533 & $1.93^{***}$ & 9060.5 \\
& 7 & 0.456 & 0.531 & $1.67^{***}$ & 8930.5 \\
& 8 & 0.470 & 0.531 & $1.39^{***}$ & 8370.0 \\
\hline
\multirow{4}{*}{Chem.} 
& 4 & 0.476 & 0.620 & $2.14^{***}$ & 9441.5 \\
& 5 & 0.500 & 0.587 & $1.45^{***}$ & 8499.0 \\
& 7 & 0.530 & 0.582 & $1.08^{***}$ & 7648.5 \\
& 8 & 0.547 & 0.591 & $0.87^{***}$ & 7276.0 \\
\hline
\end{tabular}
\caption{Mean matched factor congruence scores across instruments and factor choices. H denotes human and L denotes LLMs, and Cohen's $d$ denotes the standardized difference between the two distributions. Statistical significance is denoted by $^{***}$, for $p < .001$.}
\label{tab:factor_congruence_results}
\end{table}

\begin{figure}[t]
   \centering
   \includegraphics[width=\columnwidth]{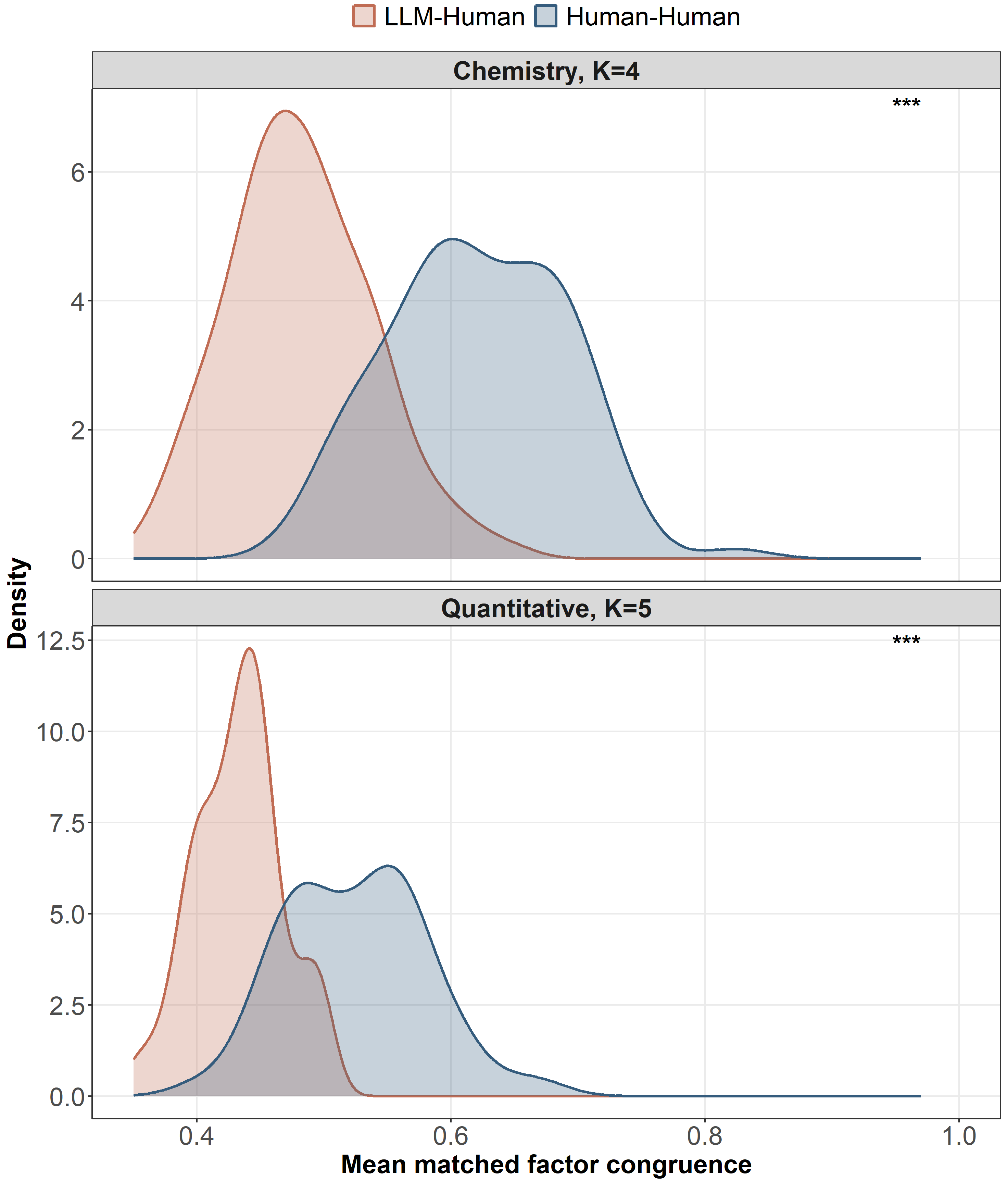}
   \caption{\textbf{Distribution of mean matched factor congruence scores}. Computed with Hungarian algorithm, for Human-Human and LLM-Human across datasets and Kaiser factor choice, on 100 resampled repetitions.}
\label{fig:factor_matching_distributions}
\end{figure}

Figure~\ref{fig:factor_matching_distributions} shows the distributions of mean matched factor congruence scores for the HH and LH comparisons across datasets and choices of the number of factors obtained for Kaiser criterion in section~\ref{sec:factor-extraction}. The HH distributions were shifted toward higher values than the LH distributions, indicating greater factor-structure similarity between independently resampled human groups than between LLMs and humans. At the same time, human-human similarity was far from perfect: the distributions were broad rather than concentrated near 1.0, indicating substantial variability across resampling iterations even within the human baseline. For the other conditions, similar results are shown in Table~\ref{tab:factor_congruence_results}. Averaged across all the tested factor solutions (4-8), mean similarity in Quantitative Reasoning was 0.535 for HH comparisons and 0.449 for LH comparisons; in Chemistry, the corresponding means were 0.595 and 0.513. One-sided Wilcoxon rank-sum tests confirmed that this difference was significant ($p < .001$) in all cases.

While we are mindful of commonly used heuristic thresholds for factor congruence (e.g., .95 for near-equivalence, .90-.94 for very high similarity, and .85-.89 for moderate similarity), the present findings are best interpreted relative to the HH baseline, because similarity within humans was itself variable across resamples. Accordingly, the main result is not perfect reproducibility in the human condition, but the consistent gap between HH and LH matching across instrument data and number of instruments. HH condition is used as an empirical baseline for expected structural agreement under repeated resampling, with the LH condition falling reliably below it.

\section{Discussion}
The main finding of this study is that both qualitative visual analysis of the factor graphs and quantitative analysis using factor congruence showed considerable differences between the latent factor structures computed through EFA for both instruments. The quantitative analysis used within-human repeated sampling as the baseline distribution and showed that LLM-human similarity remained considerably lower than human-human similarity, measured through mean matched factor congruence scores. This result was consistent across instruments and choices of the number of factors.  

Taken together, these findings answer our RQ by demonstrating that the latent structures of these evaluated case studies -- assessment instruments in science education and mathematical reasoning -- exhibit limited similarity when administered to human learners and LLMs.

Viewed through an assessment lens, these findings point to a validity-of-inference problem. The issue is not only whether LLMs obtain high scores, but whether those scores support the same construct-level interpretation that the assessment was designed to support for human respondents. In this sense, the present results broaden the validity perspective on LLM benchmarking: benchmark success may indicate that a model can produce correct answers under a given evaluation format, without establishing that the same underlying abilities are being measured. This helps explain why LLMs may perform well on a benchmark yet respond in unexpected or inconsistent ways on related tasks that appear to require the same benchmarked capabilities \citep{Kasagga2025MedicalLicensingMeta, ullman2023large}. More generally, our findings suggest that the common practice of evaluating LLMs with human-designed assessments carries a real risk of invalid interpretation if score's meaning is assumed to transfer without re-establishing validity in the new context.

If benchmark scores do not support valid construct-level inferences, then the issue extends beyond the interpretation of current results to the design of benchmarks themselves. It raises the question of how LLM benchmarks should be constructed so that the capabilities they claim to measure are clearly specified and validated on generalization that is not already available in the training data \citep{inger2025forget}. From this perspective, benchmark design must address not only task difficulty, but also the constructs being measured and whether those constructs remain interpretable across changing model architectures \citep{maimon2025iqtestllmsevaluation}. This also has important implications for educational applications built with LLMs. If students and LLMs do not share the same latent structure on a task, LLMs may be unreliable as tutors, ineffective at simulating students for various purposes, or unsuitable for assessment development.



\section{Conclusions \& Future Work}
Assessment instruments show evidence of measuring different constructs for humans and LLMs. Thus, applying assessments designed for human test-takers under the assumptions that (1) these assessments measure the same skills in LLMs, and (2) LLMs’ results on such assessments can be used to make inferences about those skills, raises validity concerns and does not adhere to educational measurement standards.

This work advances the emerging field of applying approaches from psychometrics and educational measurement to the evaluation of LLMs. Future research should broaden the scope of assessment instruments, expand the range of analytical methods employed, and incorporate human expert judgment in the interpretation of results. 





\section*{Acknowledgments}
This work was supported by the Knell Family Institute for Artificial Intelligence, Israel. The authors thank the National Institute for Testing and Evaluation for providing access to psychometric exam data.
\section*{Limitations}

We acknowledge several limitations in our data analysis that should be considered when interpreting our findings. First, using non-public datasets constrains absolute reproducibility (sec. \ref{sec: repreducibility}). We mitigate this by open-sourcing our codebase and making the complete testing instruments available to researchers upon request. Second, a key limitation regarding external validity is that the results are based on a small number of instruments and specific GenAI tools. Third, systematic biases may arise from variations in data generation procedure with LLMs such as prompting strategies, generation parameters, and User interface use. While we standardized these configurations across all LLMs, different model architectures can process identical prompts uniquely, altering response patterns. To support reproducibility and critical assessment, we have publicly released our complete codebase, prompts, and analysis scripts.

In terms of internal validity, the LLMs dataset is relatively small compared to the human sample (120 responses). We also group different LLMs together, assuming based on literature \citep{maimon2025iqtestllmsevaluation, munker-2025-fingerprinting}, that they can be treated as a single population, despite differences in their underlying architectures, which are not publicly disclosed (though they are generally assumed to be autoregressive, decoder-style transformer models).

A broader methodological limitation concerns the use of proprietary LLM systems for evaluation with non-public assessment materials. Although the evaluation instrument was private, uploading it to proprietary tools introduces a potential risk of data leakage or future contamination. Therefore, we cannot fully rule out the possibility that the instrument may become accessible to, or influence, subsequent versions of these systems.

Additionally, this study addresses one aspect of validity, namely internal structure -- other aspects, such as generalization to related tasks designed to measure the same underlying abilities, are not examined here.

\bibliography{validity}
\end{document}